\documentclass{llncs}

\usepackage{alltt}
\usepackage{framed}

\usepackage{amssymb}
\usepackage{graphics}
\usepackage{listings}
\usepackage[utf8x]{inputenc}

\usepackage{marginnote}
\usepackage{tikz}
\usetikzlibrary{shapes,arrows}
\usetikzlibrary{decorations.markings}

\newif\ifshowcomments
\showcommentstrue
\ifshowcomments
\newcommand{\mynote}[2]{\marginnote{{\bfseries\sffamily\scriptsize#1}
 {\small$\blacktriangleright$\textsf{\emph{#2}}$\blacktriangleleft$}}}
\else
\newcommand{\mynote}[2]{}
\fi

\DeclareUnicodeCharacter{8712}{{\(\in\)}}
\DeclareUnicodeCharacter{8788}{{\(:\,=\)}}
\DeclareUnicodeCharacter{8838}{{\(\subseteq\)}}
\DeclareUnicodeCharacter{8229}{{\(.\,.\)}}
\DeclareUnicodeCharacter{8696}{{\(\mkern 6mu\mapstochar \mkern -6mu\rightarrow\)}}
\DeclareUnicodeCharacter{10496}{{\(\mkern 6mu\mapstochar \mkern -6mu\twoheadrightarrow\)}}
\DeclareUnicodeCharacter{10516}{{\(\mkern 9mu\mapstochar \mkern -9mu\rightarrowtail\)}}
\DeclareUnicodeCharacter{10518}{{\(\rightarrowtail \mkern  -18mu\twoheadrightarrow\)}} 
\DeclareUnicodeCharacter{57603}{{\(\lhd\mkern-9mu-\)}} % <+
\DeclareUnicodeCharacter{9665}{{\(\triangleleft\)}}
\DeclareUnicodeCharacter{9655}{{\(\triangleright\)}}
\DeclareUnicodeCharacter{8473}{{\(\mathbb{P}\)}}
\title{An Event-B framework \\ for the validation of Event-B refinement plugins \\
               (ongoing work) 
        }
\author{J.-P. Bodeveix, M. Filali, , M.-T. Bhiri, B. Siala}
\institute{IRIT CNRS UPS Université de Toulouse \\ 
                Université de Sfax
              }
\date{October, 2016}
\usepackage{fullpage}
\begin{document}

\maketitle

\begin{abstract}
 We propose an Event-B framework for modeling the underlying theoretical foundations of Event-B.
The aim of this framework is to reuse, for Event-B itself, the  refinement development process.
This framework introduces first, a functional kernel through an Event-B context, then, it defines Event-B projects, their static and
dynamic semantics through Event-B machines. We intend to use this framework for the validation
of Event-B plugins related to distribution and for Event-B extensions related to composition and
decomposition.
\end{abstract}
\pagestyle{plain} % à commenter

\section{Introduction}

     Event-B~\cite{Abrial2010} is a method that has been proposed for building formal models together with their proofs.
As a matter of fact, it has been used for a large range of applications. Nevertheless, it
seems that, in general, it has not been applied to the field of software engineering by itself. 
In this paper, we report on an Event-B meta-framework and  two software engineering applications for which the use of the
Event-B methodology seemed to us worth to apply.  

The rest of the paper is organized as follows.  
%Section 2 gives an overview of the Event-B language. 
Section 2 outlines the main features of an Event-B framework. Section 3 discusses about two 
software applications. In conclusion, Section 4 considers some related work and sketch
future work directions.
 
%\section{A brief overview of Event-B}

%\input{framework}
\section{Towards an Event-B meta-level framework}

The proposed meta-level framework aims at validating Event-B model
transformations. We focus on transformations linked to a top-down,
refinement-based development process. Their goal is to assist the user
in producing refinements of his model through patterns parameterized
with the help of domain specific languages. Thus, a transformation pattern takes
as input an Event-B machine and some parameters. It produces either a
single machine or a set of machines. In the latter case, it is
necessary to model the project level -- not a single machine -- in order to consider the
interaction of the machines of the project. However, to make things simpler, we 
consider neither contexts, nor refinement links between
machines. 
Refinement will be taken into account at the meta level, 
each transformation producing a refinement of the project.

    \subsection{Methodology}

    We now propose a meta-level specification of an Event-B project in
    Event-B itself. The difficulty of such an exercise is to find the
    right level of abstraction and to identify which features should be
    modeled as constants and as variables. It is strongly linked with
    the objectives we have fixed. First, given the patterns we
    envision, predicates and expressions should be left as abstract as
    possible. Second, we target operations which should modify the
    project by adding new machines. Two orthogonal dynamics will thus
    be considered: project contents evolution and project operational
    semantics.  Furthermore, we try to use a refinement-based approach
    to specify the meta-level: its features will be introduced
    incrementally.

    \subsection{The global view}
\begin{figure}[hbt]
\centering
\resizebox{\linewidth}{!}{\includegraphics{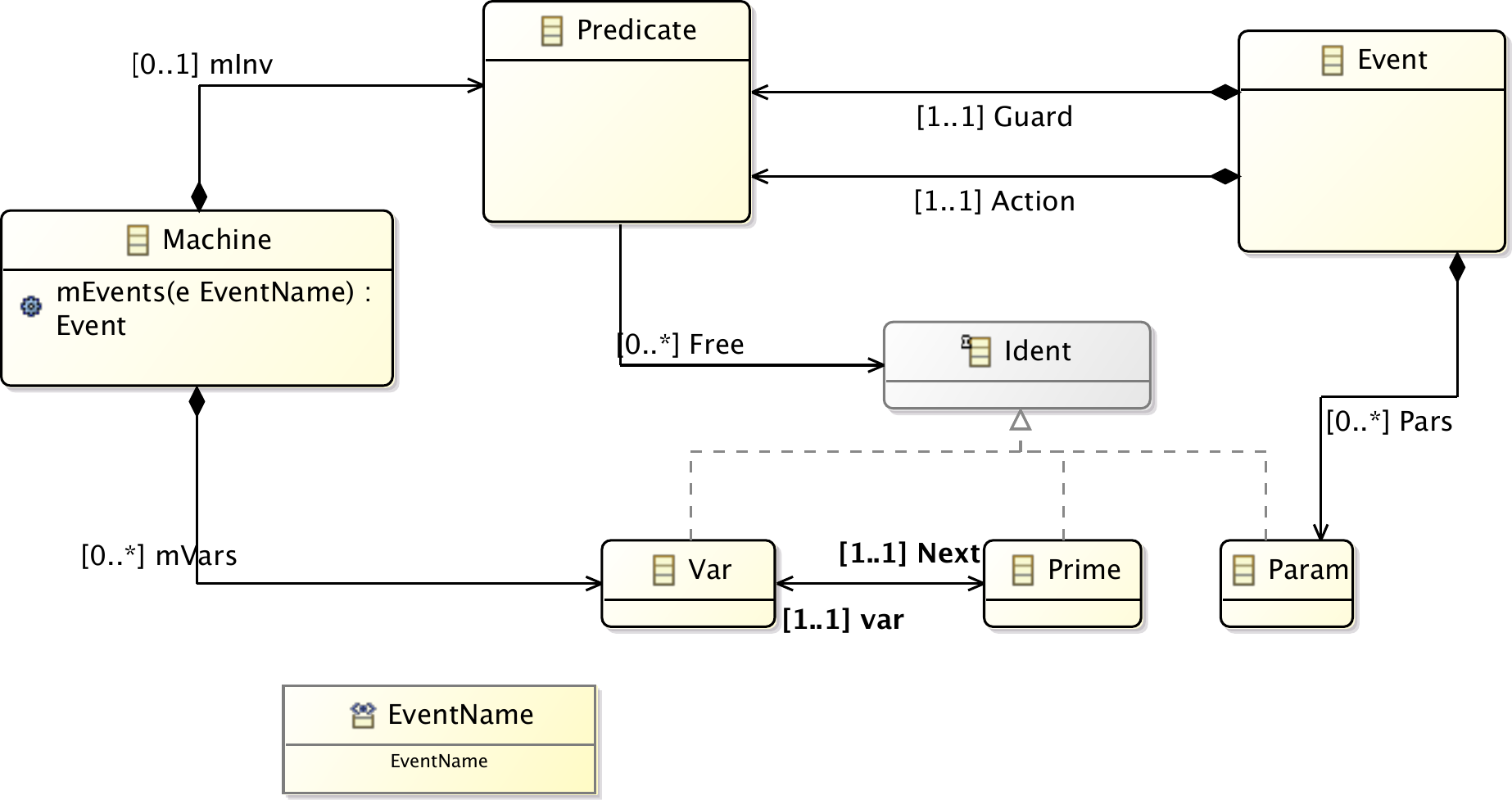}}
\caption{Event-B machines}
\label{mch}
\end{figure}

Figure \ref{mch} describes the overall structure of a machine as a
class diagram. The conversion to Event-B is performed as follows:

\begin{itemize}
  \item \texttt{Machine} is introduced as a set, with
    \texttt{Machines} being the subset of existing machines. 
  \item Machine attributes and operations can be updated and are
    defined as variables.
  \item \texttt{Predicate}, \texttt{Ident} and \texttt{EventName}. \texttt{Ident}
    is partionned into \texttt{Var}, \texttt{Prime} denoting primed
    versions of machine variables and \texttt{Param}.
  \item \texttt{Event} is modeled as a triple with three projections
    (\texttt{Pars}, \texttt{Guard} and \texttt{Action}). 
\end{itemize}

    \subsection{The functional kernel}

           The functional kernel introduces abstraction of predicates and events as Event-B contexts.
A predicate is defined as a set of abstract states. It is mainly characterized by axioms stating the existence
of the \texttt{Free} function returning the set of the free variables of a predicate and the substitution
function. With respect to our specific needs concerning decomposition/composition and distribution
we also assume the existence of a \texttt{Conjuncts} function returning a set of  predicates of which conjunct
is equivalent to the initial predicate.  For instance, the conjuncts of ``p = TRUE'' is ``\{ p = TRUE \}'' and the
conjuncts of   ``p = TRUE $\wedge$ v = 2'' is ``\{ p = TRUE, v = 2 \}''.
An excerpt of of the Predicate context is the following:

\begin{framed}
\begin{small}
\begin{alltt} 
context cPredicate extends cIdent

sets State

constants Predicate Free Subst Proj Conjuncts ...

axioms
  @Predicate_def Predicate = ℙ(State)
  @Free_ty Free ∈ Predicate → ℙ(Ident)
  @Subst_ty Subst ∈ (Ident ⇸ Ident) → (Predicate → Predicate)
  @Proj_ty Proj ∈ ℙ(Ident) → (Predicate → Predicate)
  @Conjuncts_ty Conjuncts ∈ Predicate → ℙ1(Predicate)
  @Conjuncts_ax ∀ p· p ∈ Predicate ⇒  inter(Conjuncts(p)) = p
  @Free_Conjuncts ∀ p· p ∈ Predicate ⇒  union(Free[Conjuncts(p)]) = Free(p)
 ...
\end{alltt}
\end{small}
\end{framed}

    \subsection{The Event-B project structure}

Besides contexts, Event-B projects are modelled through the following
refinement steps:

\begin{itemize}
  \item \texttt{mProject} defines the overall structure of machines
    and a project as a set of machines and provides an event to add a
    machine to a project.

  \item \texttt{static\_semantics} adds wellformedness rules
    concerning the usage of identifiers within predicates. Machine
    addition is restricted to well formed machines.

%% couper en 2???  (invariant?? et preservation de l'invariant) puis
%% semantique op (step)

 \item \texttt{dynamics} adds the invariant preservation property and
    provides a dynamic semantics to a project through the introduction
    of a state and of the \texttt{step} event defining the operational
    semantics of the project.
\end{itemize}

    \subsection{Event-B project and machines}

An Event-B project is seen as a set of machines. Each machine has
variables, an invariant and a set of events indexed by event
names. In order to make easier the meta-level reasoning, we consider
that a machine has a unique invariant and that an event has a unique
guard and a unique action (seen as a before-after predicate). These predicates will be seen as conjunctive later.

\begin{framed}
\begin{small}
\begin{alltt} 
machine mProject sees cMachine cEvent 

variables Machines mVars mInv mEvents 

invariants
  @machines_ty Machines ⊆ Machine
  @mVars_ty mVars ∈ Machines → ℙ(Var)
  @mEvents_ty mEvents ∈ Machines → (EventName ⇸ Event)
  @mInvs_ty mInv ∈ Machines → Predicate
events
  ...
end
\end{alltt}
\end{small}
\end{framed}

The \texttt{mProject} machine also provides the \texttt{new\_machine} event for adding
machines to a project. Its takes seven parameters specifying the set of
machines to be added and for each of them a set of variables, an
invariant, event names, and parameters, guard and action of each event.

   \subsection{The static semantics}

The static semantics specifies visibility constraints for variables
and parameters: 
\begin{itemize}
  \item an invariant of a machine uses variables of this machine\footnote{For the moment, we do not take into account refinements and consequently the gluing invariant.}
  \item a guard of an event can use parameters of this event and
    variables of the  machine  the event belongs to.
  \item an action of an event can use parameters of this event,
    variables of the  machine and their primed versions.
\end{itemize}

\begin{framed}
\begin{small}
\begin{alltt}
machine static_semantics refines mProject
sees cMachine

variables Machines mVars mInv mEvents 

invariants
  @mInv_ctr ∀ m · m ∈ Machines ⇒ Free(mInv(m)) ⊆ mVars(m)
  @mGuards_ctr 
     ∀ m,e· m ∈ Machines ∧ e ∈ dom(mEvents(m))
       ⇒ Free((mEvents(m);Guard)(e)) ⊆ mVars(m) ∪ (mEvents(m);Pars)(e)
  @mActions_ctr 
      ∀ m,e· m ∈ Machines ∧ e ∈ dom(mEvents(m))
       ⇒ Free((mEvents(m);Action)(e)) ⊆ mVars(m) ∪ Next[mVars(m)] ∪ (mEvents(m);Pars)(e)
\end{alltt} 
\end{small}
\end{framed}

   \subsection{The dynamic semantics}

This refinement takes into account the dynamic of a project. First,
standard proof obligations are added to express that the machine
invariant is preserved by each event. The expression of proof obligations
takes advantage of the representation of a predicate as a set:
conjunction and implication are replaced by intersection and set inclusion.
Second the operational semantics
of a project is defined through the introduction of a state for the
subset of machines considered to be active, and a \texttt{step} event modelling the evolution of the
state. The state is declared as a decomposable predicate over machine
variables. It abstracts the usual view of a state as a valuation of each state
variable. Machine invariants should be satisfied by the state.

\begin{framed}
\begin{small}
\begin{alltt}
machine dynamics refines static_semantics
sees cMachine cEvent

variables Machines mVars mInv mEvents state

invariants
  @state_ty state ∈ Machines ⇸ Decomposable     // only defined on active machines
  @state_dync ∀m· m ∈ dom(state) ⇒ state(m) ⊆ mInv(m)
  @free_state ∀m· m ∈ dom(state) ⇒ Free(state(m)) ⊆ mVars(m)
  @mInv ∀m,e· m ∈ Machines ∧ e ∈ dom(mEvents(m))
           ⇒  mInv(m) ∩ (mEvents(m);Guard)(e) ∩ (mEvents(m);Action)(e) ⊆ Subst(Next)(mInv(m))
\end{alltt}
\end{small}
\end{framed}

The \texttt{step} event makes a machine of the project advance by
updating its state. It takes as parameters a machine \texttt{m}, an
event name \texttt{e}, a predicate \texttt{p} specifying the value of
the parameters. The event guards are supposed to be satisfied by the current
state of the machine. Then its state is updated by applying the machine
action. The new state is obtained by suppressing primed in the
projection on primed variables of the conjunction of the old state,
the parameters and action predicates.

\begin{framed}
\begin{small}
\begin{alltt}
  event step
    any m e p
    where
      @m_ty m ∈ dom(state)
      @e_ty e ∈ dom(mEvents(m))
      @p p ∈ Predicate
      @f Free(p) ⊆ Param
      @g state(m) ∩ p ⊆ (mEvents(m);Guard)(e)
    then
      @a state(m) ≔ Subst(Next∼)(Proj(Next[mVars(m)])(state(m) ∩ p ∩ (mEvents(m);Action)(e)))
  end
\end{alltt}
\end{small}
\end{framed}

We also introduce an event to change the active set of machines: some
\textit{old} machines can be replaced by \textit{new} machines taken
in the pool of currently inactive machines. This event can be seen as
a hot replacement of components. It should be transparent. For this
purpose, we suppose that the conjunction of old machine states is
equal to the conjunction of new machine states. A typical application
will be to replace a compound machine by its subcomponents once it has
been split.
\section{Case studies}

  We have experimented the above meta description on two Event-B model transformations.
The first transformation deals with a safe refinement development process
for distributed applications~\cite{[SBBF16]} . This development
process proposes successive steps for splitting and scheduling complex
events. These steps are
defined by refinement patterns. They are specified through domain
specific languages.  From these specifications, two refinements were
generated. In the first phase of this work, the generated refinements
had to be verified through the Event-B framework, i.e., the Rodin
verification platform. With respect to that work, our  motivation
was to assert that the application of the proposed patterns actually
produce refinements of the source machine, so that the generated
machines are \textit{correct by construction}. Thus, it should not be
necessary to validate these refinements for each application of the
corresponding pattern.
%% C'EST LA MEME APPLICATION: les 2 transfos viennent en amont des
%% plugins de decomposition existants. Elles travaillent sur la vue centralisée
The second transformation deals with Event-B by itself. Actually, the last developments of
Event-B propose to enhance Event-B by decomposition methods. This has lead to two 
proposals: the state-based~\cite{[HA10]}  and the event-based~\cite{[SB10]}. 
Both methods have strong 
theoretical foundations. Moreover, they have been validated by significant applications and have
been both implemented by plugins available through the Rodin platform~\cite{[RCTB11]}. With respect to these
studies, our second motivation was how to \textit{manage the theoretical background} that is required
for the justification of Event-B enhancements like decomposition methods.

%       \subsection{The distribution plugin}

%        \input{shared_event}

%\input{related}

%\input{conclusion}
\section{Related Work and Conclusion}
  
    It is interesting to cite related works which have some connections
with ours.  First, Iliasov et al.~\cite{[ITLR09]} is a pioneering work
for dealing with the automation of development steps. For this
purpose, they propose the notion of refinement patterns.  Such
refinement patterns contain a syntactic description, applicability
conditions and proof obligations ensuring correctness preservation.
Unlike our approach where we stayed within an Event-B world,
\cite{[ITLR09]} adopt specific languages for representing Event-B
models and their so-called transformation rules. Last, the reuse of
the Event-B proof engine is not immediate.  Also, Cata{\~{n}}o et
al.~\cite{[CRW13]} adopt the so-called \textit{own medicine approach}
in the sense that they adopt Event-B for formalizing Event-B and JML
and the Rodin platform to discharge their proof obligations. With
respect to that our work is similar. However, their model is mainly
functional and their transformations are defined as functions. Their
correctness is stated through theorems. With respect to Event-B, we
have gone further since we have adopted a state-based approach. The
dynamic semantics as well as model transformations are defined as
events.  The correctness of the dynamic semantics and of the
transformations are obtained for free through the Event-B refinement.
Moreover, Cata{\~{n}}o et al.~\cite{[CRW13]} are concerned neither by
the validation of refinement patterns nor by the semantics of
composition.

   To conclude, Event-B proposes a refinement-based development method. In this
paper, we have studied how to support such a development method by
itself in order to formalize the underlying theoretical background:
the so-called meta level. The elaborated framework can also be used to
support Event-B enhancements as composition and decomposition methods.
As future work, we envision to broaden the coverage of our
framework. We are also interested in formalizing the links between
Event-B and temporal\cite{Hoang2016} or temporized~\cite{[GBF13]}
logics. More generally, the excplicit description of dynamic
behaviours through temporized patterns~\cite{[ADL12]} within an
Event-B framework looks challenging.
\bibliographystyle{abbrv}
\bibliography{biblio}

%\newpage

%\tableofcontents

\end{document}